# Sky Surface Brightness at Mount Graham: UBVRI Science Observations with the Large Binocular Telescope

Pedani Marco
Large Binocular Telescope Observatory, University of Arizona, 933 N. Cherry Ave, Tucson, AZ  85721 – mpedani@as.arizona.edu

**Abstract**
We present the measurements of sky surface brightness on Mount Graham International Observatory  obtained during the first binocular-mode science runs at the Large Binocular Telescope (LBT). A total of 860 images obtained on 23 moonless nights in the period Feb 2008-Jun 2008 were analyzed with our data quality assessment procedure. These data, taken at the solar minimum, show that Mt.Graham, in photometric conditions,  still has one of the darkest skies, competing with the other first-class observatories. The zenith-corrected values are 21.98, 22.81, 21.81, 20.82 and 19.78 mag arcsec$^{-2}$ in U, B, V R and I, respectively. In photometric conditions, the sky background is ~0.1 mag arcsec$^{-2}$ higher than the median when observing toward Tucson and Phoenix but it may be up to ~0.5 mag arcsec$^{-2}$ higher in non-photometric conditions. The sky at Mt.Graham is ~0.32 mag arcsec$^{-2}$ brighter at airmass ~1.4 than at zenith but no significant trend was found with the time of the night. We demonstrated the dependence of the sky background at Mt.Graham on the solar activity for the first time. In fact in 2008, at B and V bands, the sky was ~0.3 mag arcsec$^{-2}$ darker than in 1999-2002. With these results we conclude that Mt.Graham is still a first-class observing site, comparable to the darkest sites in Hawaii, Chile and Canary Islands.

## 1.   Introduction

The Mount Graham International Observatory (MGIO) is located near Safford (109º53'W, +32º42'N), Arizona, 110 Km NE of Tucson and at an altitude of 3200m. Three telescopes operate at Mt. Graham : the VATT (Vatican Advanced Technology Telescope), the Heinrich Hertz Submillimiter Telescope and the Large Binocular Telescope (LBT) with two 8.4m mirrors, which started to fully operate in binocular mode in January 2008 with its two f/1.142 prime focus cameras. Southern Arizona also hosts other observatories like Kitt Peak and Mt.Hopkins which are more affected by artificial light pollution due to the lower altitude and/or the proximity to the Tucson metropolitan area. Massey and Foltz (2000) found a increase of the zenith sky brightness at Kitt Peak and Mt. Hopkins of ~0.1-0.2 mag arcsec$^{-2}$ in the blue-optical region in the period 1988-1998. In direction of Tucson the increase was of ~0.5 mag arcsec$^{-2}$ at large zenith distances (~60°). This was the result of the increase of the population of the country land by ~27% in the same period. To mitigate this effect there are strict dark sky ordinances both in Tucson and in Safford. Thanks to its altitude, Mt. Graham is almost always above the inversion layer that forms over the desert floor, so it's less affected by dust, humidity and also by the growth of the Tucson metro area (1 million people at 2008 and mainly developing toward West). Instead it could be most exposed to the growth of Safford (21Km at AZ=50°) and Willcox (50Km at AZ=177°) in the future. Safford had a 9224 population in 2007 (-3.7% since 2000) and Wilcox had a 3787 population in 2007 (+3.7% since 2000) so the increase of the population of these two cities in the period 2000-2007 has been negligible. Nevertheless it is important to systematically monitor the sky brightness at Mt.Graham since the LBT is a big telescope with a 20+ years operating lifetime and the sky quality must be preserved also taking actions if a significant degradation should be observed in the future.

Other than human factors, the night sky brightness is also affected by several natural factors like: dust, aerosol content, zodiacal light, airglow in the upper Earth's atmosphere, solar activity. Benn & Ellison (1998) found that the sky background was ~0.25 mag brighter at an airmass (sec $z$) of 1.4 than at the zenith and that it was ~0.4mag brighter during solar maximum at La Palma. These findings were confirmed by Patat (2003) who studied the sky sbrightness at Paranal in the period 2000-2001 (Solar maximum, cycle no. 23). Interestingly both Leinert (1995) and Patat (2003) found evidence that at U-band the sky has a inverse correlation with the solar cycle (i.e. it is brighter at solar mimimum). In this work we present sky brightness measurements obtained during 23 moonless nights spread across 5 months around the solar minimum, so that, contrarily to other works that collected data across 2-3 years periods, no change of the sky brightness is expected in our data due to the changing solar cycle phase. The quality of the sky at Mt.Graham is discussed by comparing these values with the most recent sky brightness measurements obtained at the other major observatories and at Mt. Graham in the period 1999-2002 (Taylor 2004).

## 2. Observations and Data Reduction

We have obtained UBVRI sky surface brightness measurements from science images taken with the LBC prime focus cameras (Pedichini 2003) during the first fully binocular-mode runs of the LBT. The Large Binocular Camera (LBC) is a double optical imager composed by two separated large field (27 arcmin FOV) cameras, one optimized for the UBV bands and the other for the VRIz bands. The two focal plane cameras use an array of four E2V 4290 chips (4.5x2K), optimized for the maximum quantum efficiency (85%) in each channel, and a sampling of 0.22 arcseconds/pixel. Read-out noise is typically 5$e$. The data set analyzed in this work consists of 860 images taken in 23 moonless nights from February 2008 to June 2008. Images were bias subtracted and flat-field corrected using standard IRAF procedures. Since the dark current is virtually absent in the E2V 4290 chips with our typical 180-300s integration times, no such correction was applied. Only the central chip of the array of each camera (chip #2) was used to derive the sky brightness measurements. The measurement of the sky background level was obtained automatically with a modified version of the LBC data quality assessment pipeline, which uses the software Sextractor (Bertin 1996) to produce median values of the FWHM, Ellipticity, FWHM at-zenith and at V-band, for each image. For this work, we modified the output parameters of the pipeline to give also the median value of the sky background as measured by Sextractor in correspondence of any stellar-like object (typically ~100-300) detected on the chip #2. Photometric calibration was obtained using standard fields (Landolt 1992) observed during each night of the science runs. If no standard stars were observed during a particular night, the average values of the zero points and extinction from the other nights of the run were used. The final errors on the sky background measurements are 0.02-0.05 mag arcsec$^{-2}$ for all the bands. For sake of clarity it is not reported in the Figures.

If we define as photometric a night in which the zero points vary by no more than 5%, we have that 18 nights were photometric and 5 were non-photometric. Possibly a fraction of the photometric nights could have been indeed partially non-photometric since there is some overlap of the two datasets in the plots (see Fig.1).

## 3. Sky Brightness Results

In Figure 1, the UBVRI sky brightness values for each image are plotted versus the airmass. The values taken in photometric conditions (i.e. the zero points varied by no more than 5%) are plotted as filled circles, while those taken during nonphotometric conditions are plotted as open circles. Our data confirm the findings of other measurements at Mt.Graham (Taylor 2004).There is a clear difference between the sky brightness values in these two observing conditions, with the nonphotometric values being systematically brighter than the photometric ones. This is easily explained with the presence of cirrus clouds which reflect back the light pollution from the major cities. If this is the case, we should see some dependence on the azimuth with the brightest values obtained when looking toward Tucson in non photometric conditions (see Sect.3.2 for a detailed analysis). A partial overlap of the points may indicate that part of the nonphotometric night was indeed photometric. This is more evident in the B-band and V-band panels. The sky surface brightness may show high variability at even small timescales during the same night. An outstanding example is reported in Fig.15 of Patat (2003). The intensity of the airglow line [OI]6300,6364 Å increased by a factor 5.2 in two hours (from 225R to 1330R) as seen is two different spectra. Also the intensity of the airglow line [OI]5577 Å increased by a factor of 1.9 in the same period. This would imply a increase of the sky brightness at V and R-bands of up to 0.6 mag arcsec$^{-2}$. So it's not unrealistic to assume that the sky brightness can vary by an appreciable amount during the same Observing Block even in perfectly photometric nights. Nevertheless, an example of less extreme sky background fluctuations in a typical (and very clear) night at Mt.Graham (Jun 08) is visible in the U-band panel (Fig. 1). The same Observing Block was executed along a ~3.5 hrs period in which the target spanned the range 1.15-1.58 in airmass. The measurements are all included in the magnitude range U~22.0-22.1, so that the maximum fluctuation of the sky brightness at U-band was of 0.1 mag. This was our clearest and darkest night at U-band.

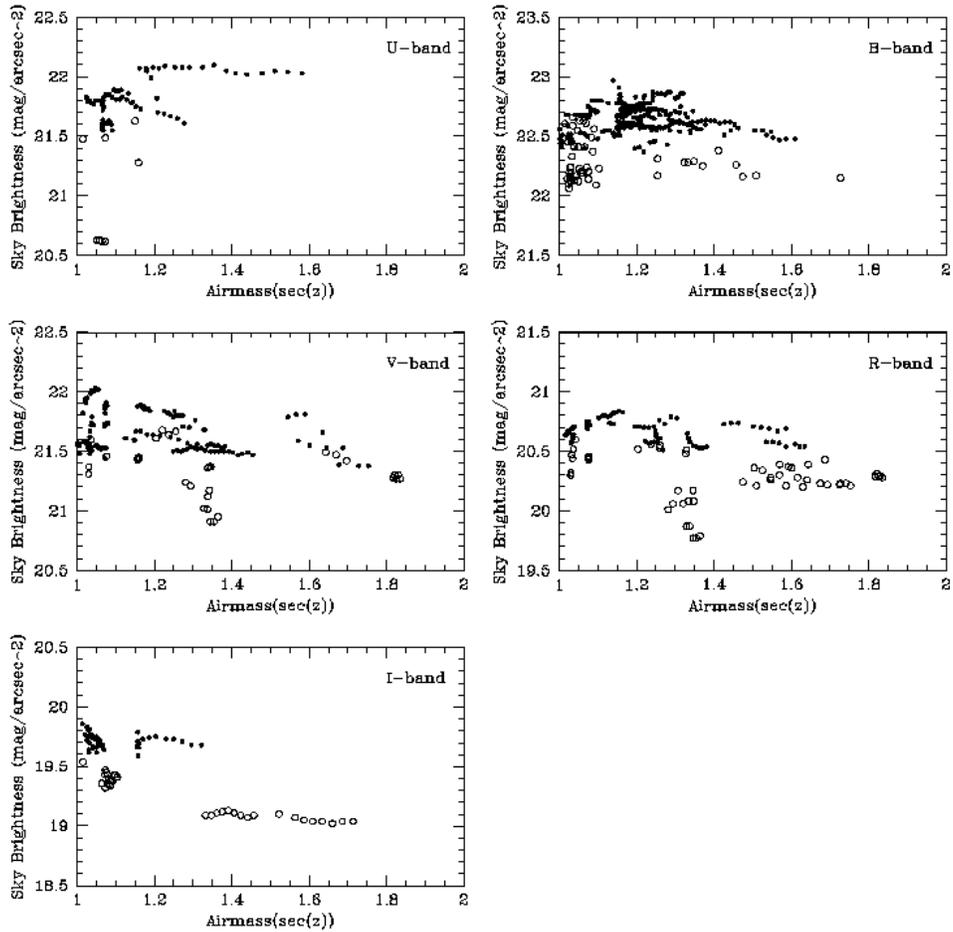

***Figure1.*** - Sky Surface Brightness measurements from the images taken at the LBT during the period Feb 2008 – Jun 2008 in U, B, V, R, I. Measurements taken in photometric and nonphotometric nights are represented by dots and open circles respectively.

### 3.1 Dependence on the Airmass

The sky brightness does increase with the airmass since the line of sight intercepts a larger path through the emitting airglow layers. When comparing the sky brightness at different sites it is important to take into account the effect of the airmass. We used the expression already reported in Patat (2003) and Sanchez (2007):

$$\Delta m = -2.5 \log [(1 - f) + fX] + k(X-1) \qquad (1)$$

where $\Delta m$ is the increase in sky brightness in magnitudes at a certain band and air mass ($X$), $f$ is the fraction of the sky brightness produced by the airglow, (1 - *f*) being the fraction produced outside the atmosphere (zodiacal light, stars), and *k* is the extinction coefficient at that band. For Mt.Graham we adopted a typical value of *f* = 0.6 as in Patat (2003) and Sanchez (2007) for the airmass correction. The mean values of the sky brightness at zenith after correction are listed in Table 1. Like in Benn & Ellison (1998) we derived the dependence of the sky brightness as a function of the airmass. We considered the photometric measurements at V, R and I-band as they have the best coverage in terms of airmass and number statistics. The U and B-bands were not considered because they have too few and/or unevenly distributed points. A linear fit gave a increase of the sky brightness at an airmass = 1.4 (zenith distance ZD = 45°) of 0.48, 0.24 and 0.23 mag at V, R and I-band respectively. The average is 0.32 mag at ZD = 45°, in good agreement with Benn & Ellison (1998) despite no corrections such as zodiacal light or ecliptic latitude were applied to our data.

### 3.2 Dependence on the Azimuth

The sky brightness increases with airmass also because of the light pollution, expecially in nonphotometric conditions and pointing toward highly populated areas. Garstang (1989) estimated the increase of the sky brightness (in magnitudes arcsec$^{-2}$) due to the light pollution from a town of P inhabitants at a distance D (Km) from the observing site for a clear atmosphere. At an airmass ~1.4, the increase of the sky brightness is:

$$\Delta m \sim ( PD^{-2.5}/70 ) \qquad (2)$$

At Mt.Graham, it would correspond to a increase of ~0.10, 0.08 and 0.06 mag arcsec$^{-2}$ when pointing toward Tucson, Phoenix and Safford respectively. However, Garstang's 1980 population models should be updated to take into account the higher lumens/per capita of the actual cities, so that those numbers may be considered as lower values. Taylor (2004) reported the predicted V-band (21.94) and B-band (22.93) sky brightness values at MGIO at zenith, solar minimum, and using 1980 population models for nearby towns and cities. It's interesting to note that our sky brightness measurements are only 0.13 mag and 0.12 mag arcsec$^{-2}$ brighter at V-band and B-band respectively. This small difference is easily explained if we take into account the increase of population since 1980. Like Taylor (2004), we analyzed the dependence of the sky brightness on the azimuth in search of any significant change in the last 6-7 years (the time elapsed from their measurements at MGIO). In Figure 2 we plotted the sky brightness measurements versus azimuth taken in photometric conditions at sec $z$ <=1.3 (filled circles) and sec $z$ >1.3 (triangles). Measurements taken in
nonphotometric conditions are plotted in Fig.3 and are represented with open circles (sec $z$ <=1.3) and asterisks (sec $z$ >1.3). Since our data were taken during a short period of time, we decided not to normalize them to the median of each observing run, as in Taylor (2004). Instead, for each band, four median values were calculated using all the available points for each observing condition (photometric, nonphotometric,  sec $z$ <=1.3 and  sec $z$ >1.3). Figure 2 shows that, in photometric conditions at U and I-band, there is no clear evidence of a dependence of the sky brightness with azimuth but this could be also affected by the poor statistics of our data. While at B-band there is no statistically significant increase of the sky brightness at 220° < az < 300° (between Tucson and Phoenix) in photometric conditions, we found that the sky background is ~0.4mag arcsec$^{-2}$ brighter in nonphotometric conditions (see Fig.3) and at high airmass.
At V and R-bands, in photometric conditions and 220° < az < 300°, the sky is ~ 0.1 mag arcsec$^{-2}$ brighter than the median. In nonphotometric conditions, however, we found it may be up to ~0.5mag arcsec$^{-2}$ brighter at V-band and ~0.3 mag arcsec$^{-2}$ at R-band.
Also, in direction of Safford, in nonphotometric conditions and high airmass, the sky background at V-and R bands is up to ~0.6 mag arcsec$^{-2}$ brighter, in good agreement with the results of Taylor (2004). This can be explained by the contribution of the three NaI lines at 5683-5688Å, 5893-5896Å and 6154-6161Å emitted by the HPS lamps which are reflected downward to the observatory by cirrus clouds. This despite Safford has a strict dark-sky ordinance. Again, no firm conclusions may be drawn in nonphotometric conditions  for the U and I-bands due to the poor statistics (see Fig.3).
We conclude that, in photometric conditions, the three major light pollution sources (Tucson, Phoenix and Safford) still have little effect on the increase of the sky background at Mt.Graham, that may be still considered one of the best available observing sites.
A more quantitative investigation of the light pollution at the MGIO and its dependence on the azimuth will be carried out as soon as the first MODS spectrograph will be installed at LBT.

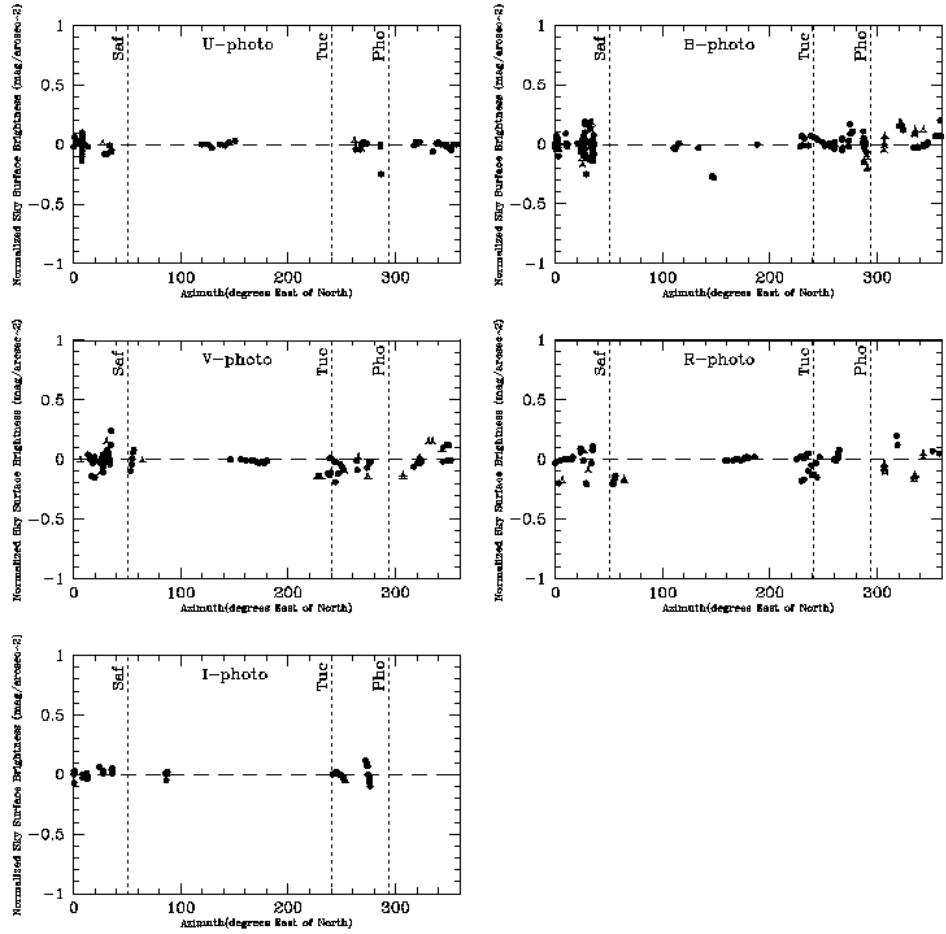

*Figure2.* - Sky Surface Brightness vs Azimuth in photometric conditions. The values are normalized to the median for each observing condition: *filled circles* for sec *z* <=1.3, *triangles* for sec *z* >1.3.
Vertical dashed line indicate the azimuth of the major cities as seen from MGIO.

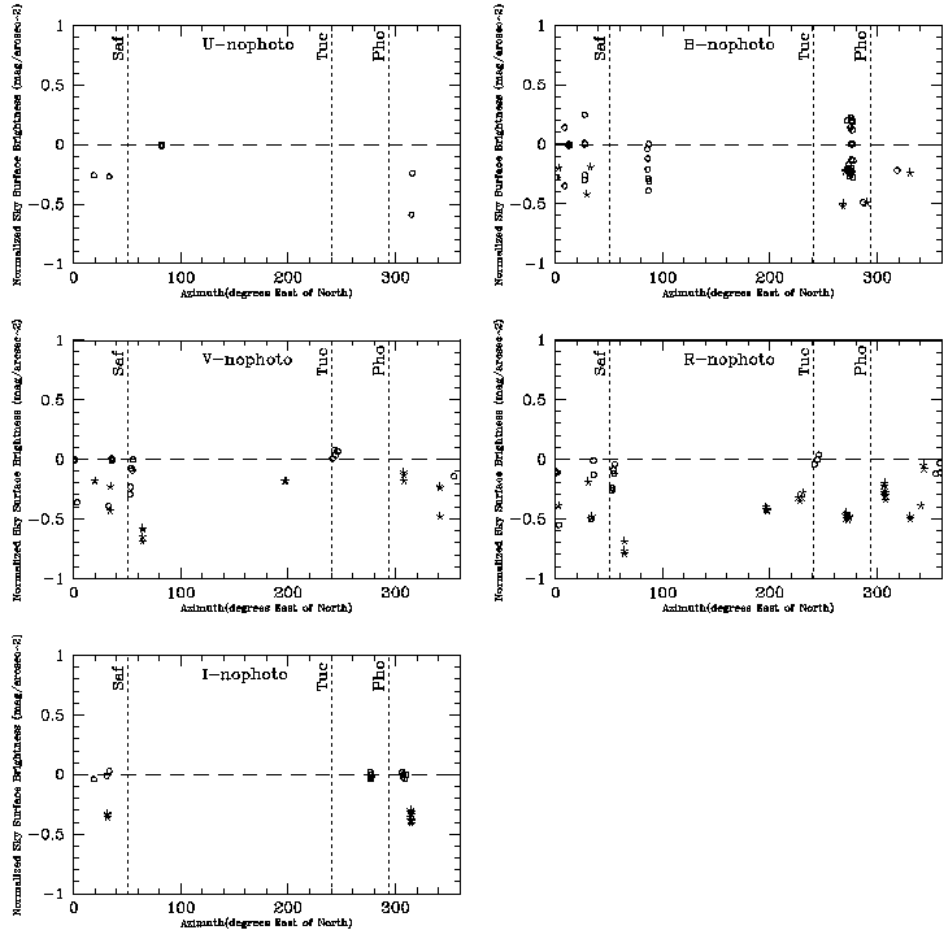

*Figure3.* - Sky Surface Brightness vs Azimuth in nonphotometric conditions. The values are normalized to the median for each observing condition: *open circles* for sec *z* <=1.3, *asterisks* for sec *z* >1.3. Vertical dashed line indicate the azimuth of the major cities as seen from MGIO.

### 3.3 Dependence on the Solar Activity

There is a general agreement that the night sky brightness is affected by the solar cycle (Walker 1988, Benn & Ellison 1998, Leinert 1995) so that this parameter must be taken into account when comparing the sky background measurements of different sites obtained at different epochs. The solar radio flux at 10.7cm wavelength is used as an indicator of the EUV solar flux in the range 100Å-650Å and the intensity of the [OI]5577 Å and [OI]6300 Å lines was found to correlate with the 10.7 cm solar flux. This correlation has a complex behavior depending on the phase of the solar cycle, time of night, time of year, geographical position and varying from cycle to cycle.
Nevertheless, Benn & Ellison (1998) found that, at solar maximum, sky is up to ~0.4-0.5 mag arcsec$^{-2}$ brighter than at solar minimum. During a typical 11 years solar cycle the 10.7cm flux is expected to vary in the range ~0.8-2.5 MJy. Since at the epoch when our data were acquired the solar 10.7cm flux was 0.70 MJy (see Tab.1), our data well represent the average observing conditions at Mt.Graham at the solar minimum and can be compared with other datasets taken at this site at different solar cycle phase. If we assume that the contribution of the light pollution at Mt. Graham did not increase appreciably in the period 2000-2008 (see Sect.1) a direct comparison with Taylor (2004) can be done.

In 2008, the sky background at Mt.Graham was as bright as in 2000 at U-band, while it was darker by ~0.28mag arcsec$^{-2}$ at B, ~0.32mag arcsec$^{-2}$ at V and marginally brighter (~0.06mag) at R. No comparison can be made at I-band since no I-band value was reported by Taylor (2004). Our data and those of Taylor (2004) were taken at a solar cycle phase of ~0.0 and ~0.70 respectively. If we assume the values of Benn&Ellison for the sky brightness variation during one solar cycle, we may estimate that, at solar maximum, the sky background at Mt.Graham, could have been ~0.1 mag arcsec$^{-2}$ brighter than that reported by Taylor (2004) (see Tab.1). Thus we derive that, at least at B and V bands, the sky at Mt.Graham is ~0.4 mag arcsec$^{-2}$ darker at solar minimum than at the solar maximum, in perfect agreement with the findings of Benn & Ellison (1998).

The case of U-band is more intriguing. Our measurements are almost identical to those Taylor (2004), despite the different solar cycle phase (see Table 1). It's interesting to note that both Patat (2003) and Leinert (1995) found a weak inverse correlation between the sky brightness at U-band and the solar cycle phase, in the sense that the U-band sky background seems to be higher at solar minimum than at solar maximum. Broadfoot & Kendall (1968) showed that, at U-band, the airglow is dominated by the O2 Herzberg bands which produce a pseudocontinuum and their intensity seems to increase with an increasing ionizing solar flux. This could at least partially explain the behavior of of the U-band sky background at solar minimum.

### 3.4 Dependence on the Night Time

Many discordant results have been presented in the past by different authors for different observing sites. Walker (1988) found a steady decrease of ~0.4mag arcsec$^{-2}$ in B and V during the first six hours after the end of twilight at San Benito Mountain. A similar amplitude variation at V-band (~0.3 mag arcsec$^{-2}$ during the first 6 hours after the end of the twilight) was reported by Krisciunas (1990) but it was not seen in B-band. On the other hand, Leinert (1995) and Mattila (1996) did not found any significant evidence of decreasing sky brightness after the end of the twilight. Also Benn & Ellison (1998) didn't find any evidence of such a variation at La Palma.

More recently, Patat (2003) did not find any evidence at least at V and R, but he found a marginal hourly rate of decrease of the sky brightness of 0.04□0.01 and 0.03□0.01 mag arcsec$^{-2}$ respectively at B and I during the first 6 hours after the end of the twilight. Nevertheless he pointed out that the fact that no steady decline was seen in V and R gave to the results at B and I bands a poorly statistically significance. This did not rule out the steep decrease as seen by other authors because many examples of that were seen in some peculiar nights at Paranal (see Fig.11 of Patat 2003). At Mt.Graham, no significant trend with the time of the night was found by Taylor (2004) at U and B. Instead, at V and R, they found a decrease of the sky brightness by 0.1 and 0.2 mag arcsec$^{-2}$ respectively at V and R during the first half of the night, followed by a slight increase by the end of the night. This is somewhat expected due to the nightly decrease of the [OI]5577 Å and [OI]6300 Å lines strength because of the recombination of ions that were excited during the day by the solar EUV radiation.

The fact that the effect found was only ~50% that observed by
Walker (1988) was attributed by Taylor (2004) to the difference in elevation between San Benito Mountain(5248 feet) and Mt.Graham (10,400 feet). Our results are plotted in Fig. 4.

Contrarily to Taylor (2004), our data do not show any significant correlation between the sky brightness and the fraction of the night for U,B,V R and I bands in photometric conditions. Since our measurements, at least at V and R bands, have twice the points plotted by Taylor (2004),
we conclude that their results could be in part due to the lower statistics. Any dependence on the solar cycle phase is ruled out since the data presented by Leinert (1995), Mattila (1986) and Patat (2003) all were taken at a phase of ~0.7-0.8, very similar to that of the run of Taylor (2004).

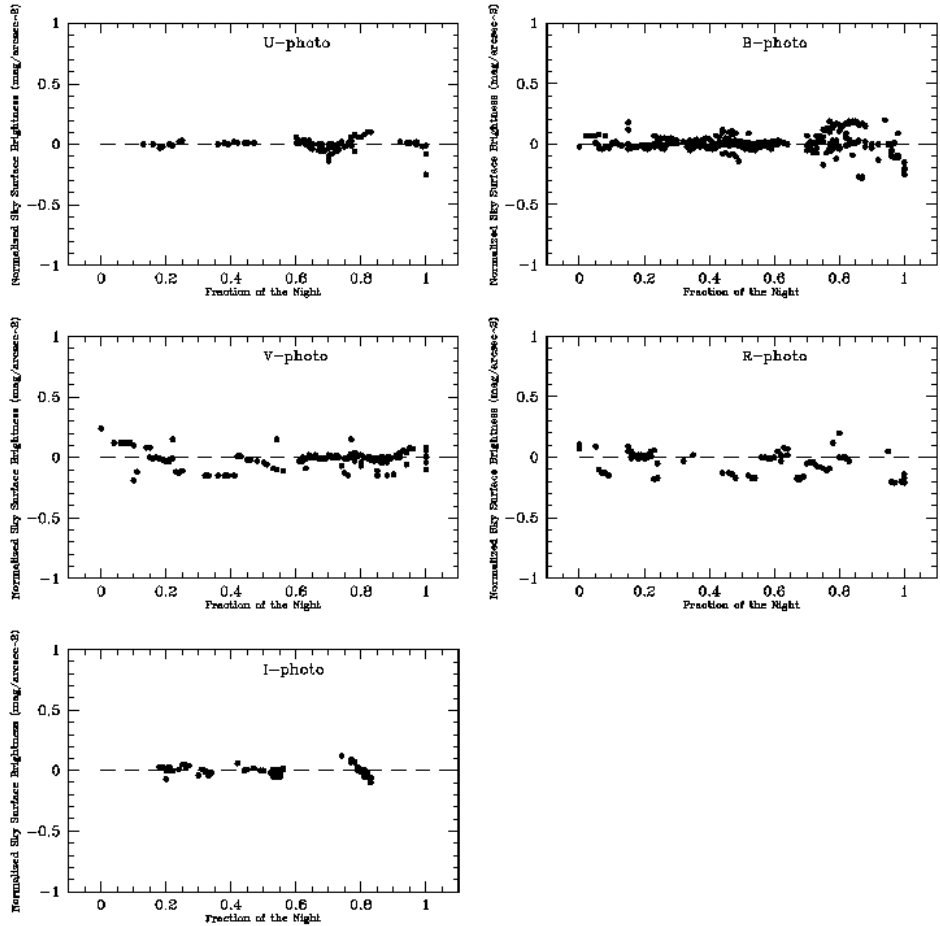

*Figure4.* - Sky brightness normalized to the median
(dashed line) for the photometric measurements. The begin
and end of the night is defined as the end and begin of
the astronomical twilight. Dusk is at fraction = 0 and
dawn is at fraction = 1.

 3.5   Comparison with Other Observatories
In Table 1 we list our zenith corrected measurements at
Mt.Graham and we compare them with those of other dark
sites from the literature. The first comparison one can
do is with the
sky brightness values at Mt.Graham for the period 1999-
2002 (Taylor 2004), even because the average solar flux
at 10.7cm at that epoch was more than twice that in 2008.
Since we can discard a significant increase of the light
pollution at Mt.Graham during the last 6-7 years, we
believe that the difference we see in the sky background
is the effect of a different solar cycle phase, as
discussed in Sect.3.3.

In 2008, the sky background at MGIO was ~0.3 mag arcsec$^{-2}$ darker at B and V, marginally (~0.06 mag) brighter at R and almost identical at U. No I-band measurements are available in Taylor (2004).

The most straightforward comparison is with Cerro Tololo, La Palma, Calar Alto and Mauna Kea, since the sky brightness at these sites was measured near the solar minimum.

From Tab. 1 we see that Mt.Graham is as dark as Cerro Tololo at U, while it is ~0.1 mag arcsec$^{-2}$ darker at B, and ~0.1 mag arcsec$^{-2}$ brighter at R and I.

According to Tab.1, Calar Alto could actually be classified as the darkest site in the world with the exception of I-band. Nevertheless, Sanchez (2007) outlined that the airmass correction formula (see Sect.3.1) does not take into account the light pollution from Almeria, a town of 190,000 inhabitants about 42km away that does not have any sky protection law like Tucson or La Palma. In fact, using Eq. (2), they estimate a $\Delta m$~0.3 mag arcsec$^{-2}$ as the contribution at zenith of the light pollution from that town. This should be compared with the contribution of the Tucson metro area at MGIO of only $\Delta m$~0.1 mag arcsec$^{-2}$ (see Sect. 3.2) obtained with the same formula.

With respect to La Palma, the sky at Mt. Graham is ~0.1 mag arcsec$^{-2}$ darker at B, ~0.1 mag arcsec$^{-2}$ brighter at V and ~0.2 mag arcsec$^{-2}$ brighter at R and I.

The sky at Mt. Graham is almost identical to that of Mauna Kea at solar minimum at B and ~0.1 mag arcsec$^{-2}$ brighter at V.

To make a direct comparison with the other sites listed in Tab.1 one should take into account the effect of the solar cycle phase which will introduce some uncertainty. The sky at Paranal, La Silla and Mt. Hopkins, adjusted to the solar minimum, could be up to ~0.2 mag arcsec$^{-2}$ darker at B and V bands than that listed in Tab.1. Thus we estimate that Mt. Graham is intrinsically brighter by ~0.1 mag arcsec$^{-2}$ at B and V than La Silla and Paranal and, respectively, ~0.1 mag arcsec$^{-2}$ and ~0.25 mag arcsec$^{-2}$ darker at B and V than Mt.Hopkins, confirming it's leadership as the best observing site in southern Arizona.

### 4.  Conclusions

We derived the average sky brightness levels at UBVRI at Mt.Graham using science images taken with the LBC prime focus cameras during the first fully binocular-mode runs of the Large Binocular Telescope. Data were taken during 23 moonless nights in the period Feb08-Jun08 for a total of 860 images. As indicated by the 10.7cm solar flux, our data were taken at the solar minimum and, for this reason, they are important to see what is the darkest sky background we can expect at the MGIO with the actual levels of light pollution.
The main results of this paper can be summarized as follows:

(1) At airmass ~1.4, the sky is ~0.32 mag arcsec$^{-2}$ brighter than at zenith at least at V,R,I, in good agreement with what found by Benn & Ellison (1988) at La Palma.

(2) The light pollution at Mt.Graham is still well under control thanks to the strict sky protection laws at Tucson and Safford. When observing toward Tucson-Phoenix in photometric conditions the sky is ~0.1 mag arcsec$^{-2}$ brighter than the median. In nonphotometric conditions, however, it may be up to ~0.5mag arcsec$^{-2}$ brighter at V-band and ~0.3 mag arcsec$^{-2}$ at R-band. Safford has more or less the same effect, in agreement with what found by Taylor (2004).

(3) The dependence of the sky brightness on the solar cycle has been clearly demonstrated by comparing our measurements to those of Taylor (2004). In 2008, the sky background at Mt.Graham was ~0.3 mag arcsec$^{-2}$ darker at B and V, marginally (~0.06 mag) brighter at R and almost identical at U. The solar flux at 10.7cm in 2008 was ~0.70 MJy and it was 1.69 MJy in 1999-2002.

(4) We did not find any relationship between the sky brightness and the fraction of the night as suggested by Taylor (2004). Our bigger sample could have played a role but some dependence on the solar cycle phase cannot be ruled out. Instead, our results are in agreement with those of other authors at La Palma, Paranal and Cerro Tololo.

(5) We conclude that Mt.Graham, thanks to its elevation and the sky protection ordinances adopted by the surrounding cities, expecially Tucson, is still a first class observing site and it's average sky background level differs by no more than ±0.2 mag arcsec$^{-2}$ from that of the best observing sites.

## 5.  References

Benn, C. R., Ellison, S. L., 1998, La Palma Technical Note, 115

Bertin, E., Arnouts, S., 1996, A&AS, 117, 393

Broadfoot, A. L., Kendall, K. R., 1968, J. Geophys. Res., 73, 426

Convington, A. E., 1969, JRASC, 63, 125

Garstang, R.H., 1989, PASP, 101, 306

Krisciunas, K., 1990, PASP, 102, 1052

Krisciunas, K., 1997, PASP, 109, 1181

Landolt, A.U., 1992, AJ, 104, 340

Leinert, Ch., Väisanen, P., Mattila, K., & Lehtinen, K. 1995, A&AS, 112, 99

Massey, P., Foltz, C.B., 2000, PASP, 112, 566

Mattila, K., Vaeisaenen, P., Appen-Schnur, G.F.O.V., 1996, PASP, 119, 153

Patat, F., 2003, A&A, 400, 1183

Pedani, M., 2004, NewA, 9, 641

Pedichini, F., Giallongo, E., Ragazzoni, R., Di Paola, A., et al., 2003, SPIE, 4841, 815

Sanchez, S.,F., Aceituno, J., Thiele, U., Perez-Ramirez, D., Alves, J., 2007, PASP, 119, 1186

Taylor, V.A., Rolf, A.J., Windhorst, R. A., 2004, PASP, 116, 762

Walker, A., 1987, NOAO Newsl., No. 10, 16

Walker, A., 1988, PASP, 100, 496

**Table 1**

Sky Surface Brightness at zenith measured at various observatories.

| Site | Year | S10.7 cm | U | B | V | R | I | Reference |
|---|---|---|---|---|---|---|---|---|
| | | MJy | mag arcsec$^{-2}$ | | | | | |
| La Silla | 1978 | 1.50 | - | 22.80 | 21.70 | 20.80 | 19.5 | Mattila (1996) |
| Cerro Tololo | 1987-88 | 0.92 | 22.00 | 22.70 | 21.80 | 20.90 | 19.90 | Walker (1987) |
| La Palma | 1994-96 | 0.80 | 22.00 | 22.70 | 21.90 | 21.00 | 20.00 | Benn & Ellison (1998) |
| Mauna Kea | 1992 | 0.71 | - | 22.84 | 21.91 | - | - | Krisciunas (1997) |
| Mauna Kea | 1996 | 2.22 | - | 22.22 | 21.29 | - | - | Krisciunas (1997) |
| Mt.Hopkins | 1998 | 1.37 | - | 22.63 | 21.46 | - | - | Massey & Foltz (2000) |
| Paranal | 2000-01 | 1.76 | 22.30 | 22.60 | 21.60 | 20.90 | 19.70 | Patat(2003) |
| La Palma* | 2003 | 1.27 | - | 22.42 | 21.63 | 20.70 | - | Pedani(2004) |
| Calar Alto | 2005-06 | 0.84 | 22.39 | 22.86 | 22.01 | 21.36 | 19.25 | Sanchez(2007) |
| Mt.Graham | 1999-02 | 1.69 | 22.00 | 22.53 | 21.49 | 20.88 | - | Taylor(2004) |
| Mt.Graham | 2008 | 0.70 | 21.98 | 22.81 | 21.81 | 20.82 | 19.78 | this work |

Note - The Penticton-Ottawa solar density flux at 2.8 GHz (10.7 cm) is also reported to show the sunspot activity at each epoch.

* From spectrophotometric data.